\documentclass[aps,reprint,groupedaddress,showpacs]{revtex4-1}
\usepackage{amsmath,amssymb}
\usepackage{graphicx}
\usepackage{epsf}
\usepackage{dcolumn}
\usepackage{bm}
\usepackage{color}

\begin{document}
\title{Photogalvanic Effect in 2D Dichalcogenides Under Double Illumination}

\author{M.V. Entin}
\email{entin@isp.nsc.ru}
\affiliation{Rzhanov Institute of Semiconductor Physics, Siberian Branch, Russian Academy of Sciences, Novosibirsk, 630090 Russia}
\affiliation{Novosibirsk State University, Novosibirsk, 630090 Russia}

\author{L.I. Magarill}
\email{levm@isp.nsc.ru}
\affiliation{Rzhanov Institute of Semiconductor Physics, Siberian Branch, Russian Academy of Sciences, Novosibirsk, 630090 Russia}
\affiliation{Novosibirsk State University, Novosibirsk, 630090 Russia}

\author{V.M. Kovalev}
\email{vadimkovalev@isp.nsc.ru}
\affiliation{Rzhanov Institute of Semiconductor Physics, Siberian Branch, Russian Academy of Sciences, Novosibirsk, 630090 Russia}
\affiliation{Novosibirsk State Technical University, Novosibirsk, 630072 Russia}
\date{\today}

\begin{abstract}
We study the photogalvanic effect caused by a simultaneous  action of circular-polarized interband and linearly-polarized intraband illuminations. It is found that, in such conditions, the steady photocurrent appears. The effect originates from the valley-selective pumping by the circular light, the trigonal asymmetry of the valleys together with the even asymmetry of the linearly-polarized light, that produces a polar in-plane asymmetry of  the electron and hole distribution functions, leading to the photocurrent. The  approach is based on the solution of the classical kinetic equation for carriers with accounting for the quantum interband excitation.
\end{abstract}


\maketitle
{\it Introduction.} The transition metal dichalcogenides (TMDs), such as $MoSe_2$,  $MoS_2$ etc. attracted much attention as real 2D materials consisting of single molecular layers \cite{TMDsBook}. These materials demonstrate unique optical and transport properties. Having the indirect-band structure in a 3D phase they become the direct band materials in a 2D phase. The band structure of these materials consists of two independent valleys coupled by the time-reversal symmetry \cite{Falko}. The large distance between valleys in the reciprocal space suppresses the intervalley scattering processes resulting in the conservation of valley quantum number. This property opens the way to use the valley quantum number as an additional to the spin degree of freedom  in modern applications for electronic devices based upon quantum laws. The new research direction, \textit{valleytronics}, attracts much attention in modern condensed matter physics \cite{rycerz, karch}.

Lately, the main attention has been paid to study the transport effects in these materials \cite{TMDs, VH1, VH2, TMDthermo}. One of the intriguing transport phenomena is the so-called valley Hall effect \cite{TMDs, VH1, VH2}. If the TMD monolayer is placed in the in-plane static electric field, the Hall current appears in each valley directed across the field. This current depends on the valley quantum number and has its opposite direction in valleys having the opposite value of valley quantum number.  The time-reversal symmetry coupling different valleys results in the vanishing of the net Hall current in equilibrium. To destroy this symmetry, the circular-polarized external electromagnetic (EM) field, having the frequency exceeding the bandgap of TMD layer, should be applied to the sample. Due to the valley selectivity of optical transitions, such EM field creates the imbalance of electron populations in different valleys resulting in the nonzero valley Hall current in the system. Usually, the circularly-polarized EM field is assumed to be weak enough and it is studied by means of the perturbation theory. Contrary, the valley Hall effect under the resonant interband excitation due to the strong EM field has been analyzed in detail \cite{we1}.

Recently \cite{we2} we have developed the quantum field theory of the coherent photogalvanic valley Hall effect in TMD materials. As for the valley Hall effect in \cite{we1}, the external EM field pumping the valleys has been considered in the nonperturbative manner. It was shown that if a sample is illuminated by two light sources having circular (with basic frequency) and linear polarizations (with double frequency), the stationary valley Hall current arises. This effect needs the intercoherence between the light sources.

The other mechanism of photocurrent which does not need two coherent light sources is considered in the present paper.
The main idea is as follows. The total electron  spectrum ${\cal E}_{\bf k}$  has its time-reversal symmetry, while the spectrum of an individual valley $\varepsilon_{\bf p}={\cal E}_{{\bf K}+{\bf p}}$ (${\bf K}$ is the position of the valley center) does not have it. At the same time,  the symmetry of the electron dispersion $\varepsilon_{\bf p}$ towards ${\bf p}\leftrightarrow -{\bf p}$, where momentum ${\bf p}$ is counted from the valley minimum, exists only approximately, near the band extrema. The lack of this symmetry means also the absence of symmetry towards the spatial reflection which is responsible for the photogalvanic effect (PGE). The valley selection by the circular-polarized light leads to a loss of the spatial symmetry and, hence, the PGE becomes permitted.

Here we will assume that the TMD layer is illuminated by two light sources. The first circularly-polarized pumping field populates the only valley, whereas the second linearly polarized probe field, creates the stationary current in the system due to intraband electron processes. Both fields are supposed to be weak, and the current arises as a second order response to the linear-polarized radiation. The frequency of circular polarized light $\Omega$ corresponds to the interband transitions, the frequency of linear-polarized light $\omega$ gets into the domain of free-electron absorption and will be assumed to be less than the characteristic energy of photoexcited carriers.

The current density in the microwave probe electric field ${\bf E}(t)={\bf E}_0\cos(\omega t)$ is described by a phenomenological expression
\begin{equation}\label{4}
j_i=\beta_{ijk}E_{0j}E_{0k}.
\end{equation}
Depending on the initial state of the dichalcogenide monolayer, one should distinguish two cases. The first one occurs if the monolayer is in the $n-$ or $p-$doped regime. In that case the circularly-polarized EM field creates the nonequilibrium electron and hole densities additional to the equilibrium ones. Formally, under the action of linearly-polarized field the PGE current in a given valley consists of the two contributions. The first one is due to equilibrium  electron ($n-$doped) or hole ($p-$doped) densities, whereas the second one is related to the photogenerated particles. The total current being summarized over the valleys does not contain the term associated with the equilibrium carriers because the latter is canceled out. Thus, in the rest of the paper we analyze the current part due to the noneqiulibrium carriers generated by the circularly-polarized EM field.

\begin{figure*}[!t]\label{fig1}
\centerline{\epsfxsize=13cm\epsfbox{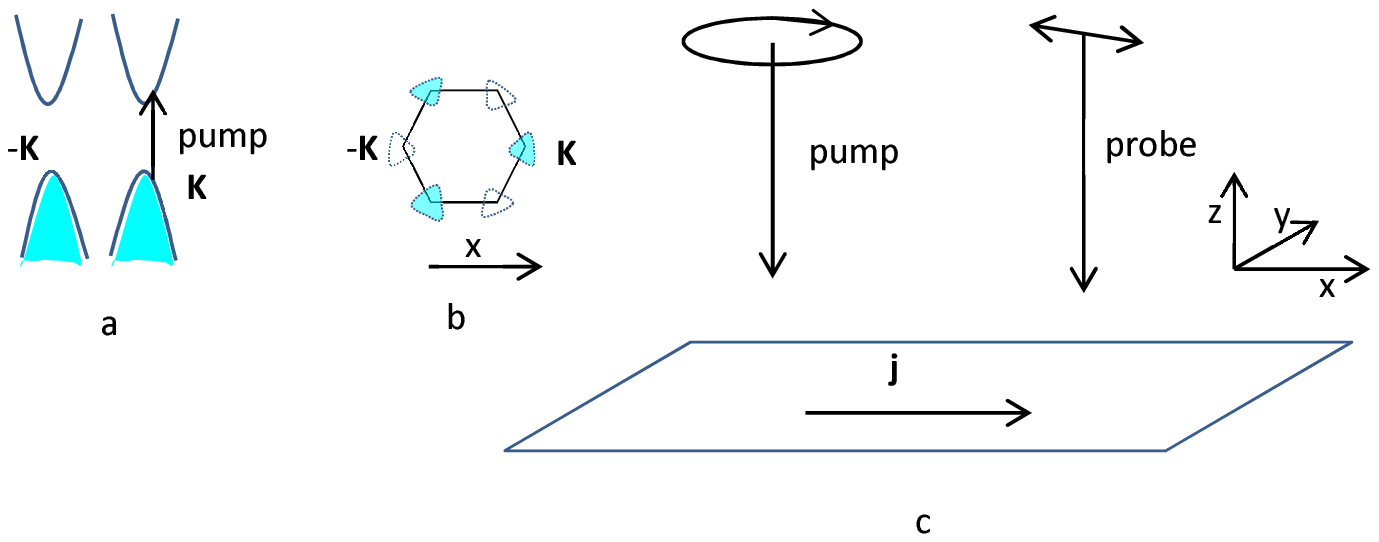}}
\caption{(Color online) a). Circular-polarized light with the right polarization selectively pumps the electrons between  the valence and conduction valleys in point ${\bf K}$. b). The Brillouin band of dichalcogenides. The ${\bf K}$ electron valley is occupied by photoexcitation, while $-{\bf K}$ remains empty. The probe linear-polarized microwave radiation together with the trigonal asymmetry of the spectrum results in the in-plane photogalvanic current. c). Sketch of predicted effect.  }
\end{figure*}

The  photogalvanic tensor $\beta_{ijk}$  in this case is proportional to $\eta\alpha J/\hbar\Omega$, where
the light, causing the interband pumping, is characterized  by the frequency $\Omega$, intensity $J$, absorption coefficient $\alpha$, and the degree of circular polarization $\eta=\eta_+-\eta_- \ (\eta_\pm$ are the fractions of a photon with right or left polarizations).

The symmetry of point ${\bf K}$ in a dichalcogenide is $C_{3v}$. Taking into account the fact that the linear tensor $\beta_{ijk}$ originates from the spectrum warping, we find the non-zero components of  $\beta_{ijk}$: $\beta_{yxy}=\beta_{yyx}=\beta_{xyy}=-\beta_{xxx}=-\beta$. Thus, it is enough to calculate the component $\beta$ only.

For simplicity, below we neglect the bands spin splitting. The spin is a robust quantum number for interband transitions. The theory will be developed for a single spin projection. The generalization to both spin projections is reduced to collecting  the corresponding contributions to the current.

The other simplification is the neglection of weak non-selectivity of interband transitions caused by the interband matrix element dependence on momentum ${\bf p}$. This assumption is valid if $\Omega$ only slightly exceeds the badgap $\Delta$, $\Omega-\Delta\ll\Delta$ (we set $\hbar=1$). The generalization to the finite $\Omega-\Delta$ can be done by means of introducing $\gamma$, the selectivity of excitation into valley $\bf K$.  In the two-band Dirac Hamiltonian \cite{Falko,TMDs,we1} disregarding the possible spin splitting of the valence band, $\gamma=2\Delta\Omega/(\Delta^2+\Omega^2)$. At the absorption threshold $\Omega=\Delta$, $\gamma=1$, and at $\Omega\gg \Delta$, $\gamma\to 0$.

We, initially, discuss the valley selective pumping and then present the PGE theory based on the Boltzmann equation accounting for the intraband impurity scattering. This approach is justified by the assumption that the intravalley relaxation is much quicker than the recombination and intervalley ones, allowing for  the separation of the process into two stages. The low symmetry of point $\bf K$ is reflected in the electron spectrum $\varepsilon_{\bf p}=\epsilon_p+w_{\bf p}$, where  $\epsilon_p=p^2/2m$ is the electron energy, $w_{\bf p}=C_3(p_x^3-3p_xp_y^2)=C_3p^3\cos(3\phi_{\bf p})$ is the trigonal warping correction, ${\bf p}=p(\cos\phi_{\bf p},\sin\phi_{\bf p})$ (this choice corresponds  to the reflection symmetry relative to the axis $x$). It is known that the asymmetry of the particle dispersion leads to the second harmonic generation \cite{GT}, purely valley currents \cite{MEGT} and alignment \cite{Portnoi} of the photoexcited carriers in gapless materials (graphene). In the present paper just the quantity  $w_{\bf p}$ is responsible for PGE.

The field ${\bf E}(t)$ interaction with electrons is described in the framework of classical Boltzmann equation.  The scattering by neutral and charged impurities will be considered.

{\it Valley pumping.} The possibility of PGE is determined  by the disbalance between the valley populations $n_{\bf K}-n_{-\bf K}$. The valley concentrations $n_{\pm\bf K}$ are controlled by the circular polarized interband radiation. In the equilibrium $n_{\bf K}=n_{-\bf K}$; the same is true if two circular components are mixed at an equal proportion. The stationary values of  $n_{\bf K}$ are determined by the balance between generation and recombination or intervalley scattering processes. We will assume that the probability of these processes is much less than the intravalley scattering. In this case, the intravalley equilibrium is established, while the disbalance between valleys remains. The balance is described by the equation
\begin{gather}\label{pop1}
\frac{\delta n_{-\bf K}-\delta n_{\bf K}}{\tau_v}-\frac{\delta n_{\bf K}}{\tau_r}+g_{\bf K}=0,
\end{gather}
where $\delta n_{\bf K}$ is a concentration of photoexcited carriers, $\tau_v$, $\tau_r$ are intervalley and recombination times, $g_{\bf K}$ is a photogeneration rate, accordingly. The quantity $g_{\bf K}$ is determined by
\begin{gather}\label{gK}
g_{\pm{\bf K}}=\alpha J\eta_\pm\gamma/\hbar\Omega.
\end{gather}
From Eq.(\ref{pop1}) we find
\begin{gather}\label{pop2}
\delta n_{\bf K}-\delta n_{-{\bf K}}=\tau_0\alpha J\eta\gamma/\hbar\Omega,
\end{gather}
where $1/\tau_0=1/\tau_r+2/\tau_v$. If there are finite valley populations at equilibrium ($n,p-$doped regime), the relaxation times do not depend on the external illumination, whereas, in the absence of carriers at equilibrium, the relaxation time depends on the $\delta n_{\pm\bf K}$ and, as a result on pumping field intensity $J$. Note, that the values of $\tau_0$  for electrons and holes (and, consequently, their concentrations) can differ.

{\it Calculation of PGE coefficient.} Our analysis of PGE coefficient $\beta$ is based on the classical kinetic equation approximation.
For simplicity, we consider one type of carriers, namely electrons, keeping in mind that the total current is determined by the summation of contributions from different types of carriers and valleys. Let us have the valley ${\bf K}$ to be populated, while  $-{\bf K}$ is empty. Note here that the case when the pumping field has its arbitrary polarization is described by the same final equations with the additional factor $\eta$ reflecting the circular polarization degree. In specific cases of positive or negative circular polarization $\eta=\pm 1$.

The classical kinetic equation for  the distribution function $\mathcal{F}({\bf p},t)$ reads
\begin{equation}\label{5}
\partial_t\mathcal{F}+e{\bf E}\nabla_{\bf p}\mathcal{F}=\hat{I}\mathcal{F}.
\end{equation}
The quantity $\hat{I}$ represents the collision operator with impurities.  In the Born approximation
\begin{equation}\label{7.3.3}
\hat{I}\mathcal{F}({\bf p})=2\pi n_i\int \frac{d^2 p'}{4\pi^2}|V({\bf p}-{\bf p}')|^2\delta(\varepsilon_{\bf p}-\varepsilon_{\bf p'})(\mathcal{F}({\bf p})-\mathcal{F}({\bf p}')).
\end{equation}
Here   $V({\bf p})$ is the Fourier transform of the impurity potential, $n_i$ is  the impurity concentration. Independently from the spectrum asymmetry, the collision operator vanishes if  $\mathcal{F}({\bf p})$ depends only on the electron energy (in particular, if it is the equilibrium distribution function). The collision operator bears the valley asymmetry via the electron spectrum and can be separated into two parts, $\hat{I}=\hat{I}_++\hat{I}_-$ . The first one,  $\hat{I}_+$ contains the isotropic part of electron spectrum, $\epsilon_p
$, whereas   $\hat{I}_-$ is determined by the spectrum asymmetry. Assuming $\hat{I}_-\ll\hat{I}_+$ and expanding Eq.(\ref{7.3.3}) with respect to $w_{\bf p}$, we find
\begin{eqnarray}\label{10}\nonumber
&&\hat{I}_-\mathcal{F}({\bf p})=2\pi n_i\int \frac{d^2 p'}{4\pi^2}|V({\bf p}-{\bf p}')|^2\times\\
&&\delta'(\epsilon_p- \epsilon_{p'})(w_{\bf p}-w_{\bf p'})(\mathcal{F}({\bf p})-\mathcal{F}({\bf p}')).
\end{eqnarray}

The solution of Eq.(\ref{5}) is executed by the expansion in electric field powers. The stationary correction appears in the second-order with respect to the electric field. It is given by
 \begin{equation}\label{6}
\overline{\mathcal{F}^{(2)}}=\frac{e^2}{2}\mbox{Re}\left(\hat{I}^{-1}{\bf E}_0\nabla_{\bf p}\left[(i\omega+\hat{I})^{-1}{\bf E}_0\nabla_{\bf p}f({\bf p})\right]\right).
\end{equation}
Here overline stands for the time averaging, \ $f(\textbf{p})$ is the stationary distribution function of photogenerated carriers depending on the particle spectrum warping. This function is produced by the pumping. It is assumed that, due to long intervalley and recombination times, $f(\textbf{p})$ is quasiequilibrium function within the pumped valley. At the same time, this function is anisotropic in the momentum space due to the spectrum warping.

The contribution to the stationary current density of electrons is expressed via $\overline{\mathcal{F}^{(2)}}(\textbf{p})$ as
\begin{equation}\label{7}
{\bf j}=e\int( \nabla_{\bf p}\varepsilon_{\bf p})\overline{\mathcal{F}^{(2)}}(\textbf{p})\frac{d^2 p}{4\pi^2}.
\end{equation}

The photogalvanic current Eq.(\ref{7}) arises due to the asymmetry of the spectrum directly and via the collision operator. Assuming the smallness of $w_{\bf p}$  we expand the current with respect to $C_3$. The expansion touches on the velocity operator $\nabla_{\bf p}\varepsilon_{\bf p}$, stationary distribution function $f({\bf p})$, and the collision operator $\hat{I}$. Thus, we have
\begin{gather}\label{7.1}
\nabla_{\bf p}\varepsilon_{\bf p}=\frac{\textbf{p}}{m}+\nabla_{\bf p}w_{\bf p}, \ \
f({\bf p})\approx f_0(\textbf{p})+w_\textbf{p}\frac{\partial f_0(\textbf{p})}{\partial\varepsilon_\textbf{p}},
\end{gather}
where $f_0(\textbf{p})$ is the stationary particle distribution function taken at $w_\textbf{p}=0$. The collision integral can be also expanded with respect to the asymmetric part
\begin{gather}\label{7.2}
\hat{I}^{-1}=\hat{I}_+^{-1}-\hat{I}_+^{-1}\hat{I}_-\hat{I}_+^{-1}+...,\\\nonumber
\frac{1}{i\omega+\hat{I}}=\frac{1}{i\omega+\hat{I}_+}-\frac{1}{i\omega+\hat{I}_+}\hat{I}_-\frac{1}{i\omega+\hat{I}_+}...
\end{gather}
Thus, we have three contributions to the current which comes from the velocity operator, stationary distribution function and the collision operator, respectively.

At $C_3=0$ the system is isotropic in the $(x,y)$ plane. In this case, the action of $\hat{I}_+$ onto the $M$-th angular harmonics of distribution function $\mathcal{F}({\bf p})\propto e^{iM\phi_{\bf p}}$ reduces  to the multiplication by the corresponding relaxation rate:
  \begin{equation}\label{8}
\hat{I}_+\mathcal{F}({\bf p})=-\mathcal{F}({\bf p})/\tau_M,\end{equation}
where
 \begin{equation}\label{tauM}
\frac{1}{\tau_M}=2\pi mn_i\int |V({\bf p}-{\bf p}')|^2(1-\cos (M\theta))d\theta.
\end{equation}
Here $p=p'$, $\theta$ is an angle between ${\bf p}$ and ${\bf p}'$. For the further calculation we use the identity
\begin{gather}\label{8.1.2}
\frac{1}{\alpha+\hat{I}_+}[g(p)\cos (M\varphi_\textbf{p})]=\frac{g(p)\cos (M\varphi_\textbf{p})}{\alpha-1/\tau_M}
\end{gather}
valid for any function $g(p)$.

In the following section we calculate these contributions assuming the simplified $\tau-$constant approximation of the isotropic collision integral $\hat{I}_+=-1/\tau$ which corresponds to the electron scattering by short-range impurities. This approximation is also valid for Coulomb scattering when the screening length becomes less than the electron wavelength. Although this condition can be hardly fulfilled if impurities are screened by the same carriers, it is valid if there is an additional screening mechanism, for example due to the presence of gate electrode.  The generalization to the case of unscreened Coulomb impurities will be given in a further section.

{\it Short-range impurities. } The contribution to the current which corresponds to the correction $\nabla_{\bf p}w_{\bf p}$ is given by
\begin{gather}\label{7.1.1}
\textbf{j}_1=-\frac{e^3\tau^2}{2(1+\omega^2\tau^2)}\int\frac{(\nabla_{\bf p}w_{\bf p})d^2p}{(2\pi)^2}(\textbf{E}_0\cdot\nabla_\textbf{p})^2f_0(\textbf{p})
\end{gather}
It can be readily shown that the other contributions to the PGE current vanish.

For the scattering on short-range impurities, $1/\tau=mn_iU_0^2$ does not depend on the electron energy. Substituting $w_{\bf p}$ in Eq.(\ref{7.1.1}), we find
\begin{gather}\label{7.1.2}
j_{1x}=\frac{e^3\tau^2E^2_{0x}3C_3}{2(1+\omega^2\tau^2)}\int\frac{(p_x^2-p_y^2)d^2p}{(2\pi)^2}\frac{\partial^2f_0(\textbf{p})}{\partial p_x^2}.
\end{gather}

The last integral can be directly expressed via the total density of photo-generated electrons in a given valley
$$n_{\bf K}\equiv n=\int\frac{d^2p}{(2\pi)^2}f_0(\textbf{p}).$$
Finally, the total  PGE coefficient including the summation over carrier kinds is
\begin{gather}\label{7.1.33}
\beta=3e^3\eta\gamma \left[\frac{n_e\tau_e^2C_{3e}}{1+\omega^2\tau_e^2}-\frac{n_h\tau_h^2 C_{3h}}{1+\omega^2\tau_h^2}\right],
\end{gather}
where the subscripts $e$ and $h$ indicate  carrier types.

It is interesting to note that this result is analogues to the photon drag effect \cite{ivch} with the change of $3C_3\eta\gamma\rightarrow 2q/(\omega m^2)$, where $q,\omega$ are the in-plane component of photon wavevector and its frequency.

{\it Coulomb impurities.} Here the scattering caused by the non-screened charge-impurities with concentration $n_i$ is studied. The Fourier transform of the impurity potential is $V({\bf q})=2\pi e^2/(\kappa q)$ ($\kappa$ is the dielectric constant).
The quantity $\tau_M$ is
 \begin{equation}\label{tauM}
\frac{1}{\tau_M}=\frac{\pi me^4n_i}{\kappa^2p^2}\int d\phi\frac{1-\cos (M\phi)}{1-\cos \phi}=\frac{\pi^2 e^4n_i}{\kappa^2\epsilon_p}|M|.
\end{equation}

 The dependence of the Coulomb scattering times  on the electron energy (see Eq.(\ref{tauM})) brings additional contributions to the current.  Eq.(\ref{10}) converts to
\begin{eqnarray}\label{11}\nonumber
&&\hat{I}_-\mathcal{F}({\bf p})=\frac{m^2 n_ie^4C_3}{2\kappa^2p}\int\limits_0^{2\pi} d\phi\frac{d}{dp'}\Bigg[(\mathcal{F}({\bf p})-\mathcal{F}({\bf p}'))\\&&\times\frac{p^3\cos(3\phi)-p'^3\cos(3\phi')}{p'^2+p^2-2pp'\cos(\phi-\phi')}\Bigg]_{p'=p}.
\end{eqnarray}

Further calculations are performed with the use of angular harmonics and Eqs. (\ref{8.1.2}, \ref{tauM}, \ref{11}). We found the current for the Boltzmann distribution function of photoexcited carriers.
The resulting PGE coefficient reads
\begin{eqnarray}\label{finCoulomb}
\beta=16e^3\eta \gamma\tau^2(n_hC_{3h}-n_eC_{3e})F(\omega\tau),\\
F(y)=\frac{1}{32}\int\limits_0^\infty  \frac{x^2e^{-x}dx}{4+x^2y^2}\times\nonumber\\ \left[4x^2+\frac{2x+10-20x^2y^2+x^3y^2-x^5y^4}{(1+x^2y^2)^2}\right] \label{F},
\end{eqnarray}
where
\begin{gather}\label{tau}
\tau=\frac{\int_0^\infty\tau_1e^{-\epsilon/T}d\epsilon}{\int_{0}^{\infty}e^{-\epsilon/T}d\epsilon}=\frac{T\kappa^2}{2\pi^2 e^4n_i}.
\end{gather}
The function $F(y)$ having the asymptotic behaviour
\begin{gather}\label{FFF}
F(y)\approx 1-\frac{135}{8}y^2~~~\mbox{if}~~~y\ll 1,\\\nonumber
F(y)\approx \frac{7}{32y^2}~~~\mbox{if}~~~y\gg 1,
\end{gather}
is presented in Fig.2. For  unscreened Coulomb impurities scattering the relaxation times are identical for electrons and holes, Eq.(\ref{beta}). In that case the total PGE of electrons and holes is determined by the same function $F(y)$ in Eq.(\ref{finCoulomb}) with the renormalized coefficient $\tau^2 nC_3\rightarrow \tau_e^2n_eC_{3e}-\tau_h^2n_hC_{3h}$.

In a simple  situation $n_e= n_h$ for interband transitions. In a two-band model $C_{3e}=C_{3h}$. The Coulomb scattering also does not depend on the charge sign, and $\tau_e=\tau_h$. In such case the PGE current should vanish. Nevertheless, $C_{3e}\neq C_{3h}$ due to the influence of other bands on the electron spectrum (see, e.g. \cite{Falko}). Besides, $n_e\neq n_h$ if these quantities are controlled by different mechanisms of carrier capture.

\begin{figure}[h]\label{fig2}
\centerline{\epsfxsize=7cm\epsfbox{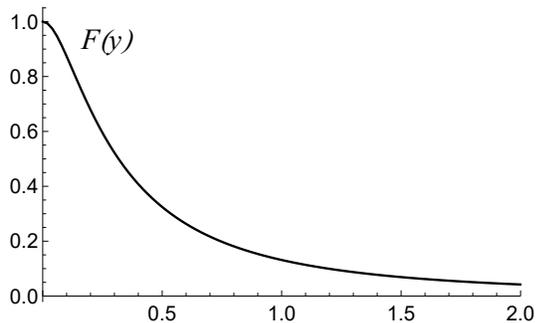}}
\caption{Function $F(y)$ according to Eq.(\ref{F}). }
\end{figure}
As seen in Fig.2 (unlike the case of short-range impurity scattering) the  PGE experiences a non-uniform scaled drop with the frequency.  This is the consequence of concurrence between different contributions to the PGE. One of them appears as a result of asymmetric scattering of the stationary even corrections to the distribution function, producing the odd stationary distribution function responsible for the current. The other results from the high-frequency odd correction to the distribution function, which, after the asymmetric scattering, gives rise to the even high-frequency function and,  after an additional action of microwave field, converts to the stationary odd function. The  corrections to the current are controlled  by dimensionless parameters $\omega\tau_1$ or $\omega\tau_2$, and that results in their different relative strength in different frequency domains.

Let us estimate the  effect in $MoS_2$ by means of Eq. (\ref{finCoulomb}). The parameters $C_{3e}$  and $C_{3h}$ have values $C_{3h}=-5.71\,eV{\AA},~C_{3e}=-3.49\,eV{\AA}$ \cite{Falko}.  Choosing the values of other necessary parameters $\tau_e=\tau_h=10^{-10}s$, $n_e=n_h=10^{10}\,cm^{-2}$ and taking $\eta=\gamma=1$, we find $\beta=2.3\mu A\cdot cm/V^2$ at $\omega=0$.

{\it Conclusions and discussions.} The effect studied here  differs from the known photogalvanic effect (see \cite{GlazovGanichev} and references therein) by its
origin from the spectrum asymmetry rather than the asymmetry of
interaction potential or the crystal-induced Bloch wave function asymmetry. The circular polarized light
causes the selective population of valleys; the linear-polarized
microwave illumination converts the
trigonal asymmetry of the spectrum in a single valley to  the polar
asymmetry of the photocurrent. The  valley population lives during a relatively
long intervalley time, as compared with the momentum relaxation time (responsible for usual PGE). This circumstance amplifies the effect.

The photocurrent is determined by both electrons and holes. As seen in Eq.(\ref{7.1.33}), the e-h asymmetry (in particular, difference between $\tau_e$ and $\tau_h$),  together with a different direction of partial currents,  results in the complex  frequency dependence of the photocurrent, up to its alternating sign.

It is desirable to compare the effect predicted here to the other PGE. The usual PGE appears in the second order in the external
alternating electric field. It requires the medium asymmetry. The coherent PGE \cite{we2} appears, at least, in the third order of
electric field, does not require the asymmetry of the media, but needs the
intercoherence of two light sources with the first and second harmonics.

As
compared to the coherent PGE, the effect considered here has the forth
order in the external field, which can be produced by two independent light sources. Hence, the
inter-coherence of two light sources is not assumed.

The considered effect depends on different parameters describing the system,
namely, the asymmetry of valley spectrum, intervalley scattering time, momentum
relaxation time and valley pumping selectivity. All these factors contribute to
the effect and, hence, can be studied by the effect measurements.

{\it Acknowledgments.} This research was supported by the RSF grant No 17-12-01039.

\end{document}